\documentclass[12pt,a4paper]{article}
\newcommand{\lesssim}{ \mathop{}_{\textstyle \sim}^{\textstyle <} }

\newcommand{\GeV}{\mbox{\rm GeV}}
\newcommand{\eV}{\mbox{\rm eV}}

\def\vev#1{\left\langle #1\right\rangle}

\def\abs#1{\left| #1\right|} 

\newcommand{\eqn}[1]{&\hspace{-0.5em}#1\hspace{-0.5em}&}
\begin{document}
\begin{titlepage}

~

%%%%%% Preprint #
\begin{flushright}
UNIL-IPT-03-2
\end{flushright}

\vskip 1.5cm

%%%%%% Title
\begin{center}
{\large {\bf Lopsided Mass Matrices and Leptogenesis in SO(10) GUT}}

\vskip 1.2cm

%%%%%% Authors
{T. Asaka}

\vskip 0.4cm
            
%%%%%% Addresses
{\it Institute of Theoretical Physics, University of Lausanne,\\
CH-1015 Lausanne, Switzerland}

\vskip 0.2cm

%%%%%% Date
%(\today)
(April 2003)

\vskip 2cm

%%%%%% Abstract
%\maketitle
\end{center}
\begin{abstract}
  Lopsided structure in mass matrices of down quarks and leptons gives 
  a simple explanation for the observed large angles of neutrino mixings.
  We realize such mass matrices by the Froggatt-Nielsen mechanism
  in the framework of supersymmetric SO(10) grand unified theory (GUT).
  It is shown that the model can reproduce the successful mass
  matrices which have been obtained in SU(5) models.
  Cosmological implication of the model is also discussed.
  We show that the hybrid inflation occurs naturally in the model
  and it offers non-thermal leptogenesis by decays of the next-to-lightest
  right-handed neutrinos.  The present baryon asymmetry is explained 
  by just the oscillation mass scale in the atmospheric neutrinos.
\end{abstract}
%%%%%% Abstract
\end{titlepage}
%%%%%%%%%%%%%%%%%%%%%%%%%%%%%%%%%%%%%%%%%%%%%%%%%%%%%%%%%%%%%%%%%%%%
%%%%%%%%%%%%%%%%%%%%%%%%%%%%%%%%%%%%%%%%%%%%%%%%%%%%%%%%%%%%%%%%%%%%
\newpage
\renewcommand{\thefootnote}{\arabic{footnote}}
\setcounter{footnote}{0} \setcounter{page}{1}
%%%%%%%%%%%%%%%%%%%%%%%%%%%%%%%%%%%%%%%%%%%%%%%%%%%%%%%%%%%%%%%%%%%%
%%%%%%%%%%%%%%%%%%%%%%%%%%%%%%%%%%%%%%%%%%%%%%%%%%%%%%%%%%%%%%%%%%%%
Recent neutrino experiments have provided quite convincing evidence
for neutrino masses and their flavour mixings. The atmospheric
neutrino anomaly is explained by the $\nu_\mu$-$\nu_\tau$ oscillations
with $\delta m^2_{\rm atm} \simeq 2.5 \times 10^{-3} ~\eV^2$ and
$\sin^22\theta_{\rm atm} \simeq 1.0$~\cite{Fukuda:2000np}, while the
solar neutrino problem is now solved by the so-called ``large-mixing
angle solution'' with $\delta m^2_{\rm sol} \sim 7 \times 10^{-5}
~\eV^2$ and $\tan^2 \theta_{\rm sol} \sim 0.5$~\cite{Fukuda:2001nk}.
These developments give us an important clue to physics beyond the
Standard Model.  

The most natural extension is probably to introduce right-handed
neutrinos with superheavy Majorana masses, since they induce the
observed suppressed masses of neutrinos through the seesaw
mechanism~\cite{SeeSaw}.  Further, non-equilibrium decays of
right-handed neutrinos in the early Universe offer one natural way to
generate the present baryon asymmetry~\cite{Fukugita:1986hr}.  Then,
grand unified theory (GUT) based on SO(10) gauge group~\cite{SO10} is
particularly attractive, since all quarks and leptons of each family
are unified into a single spinor representation together with the
right-handed neutrino.  For unification of gauge couplings and
stabilization of the gauge hierarchy we had better incorporate
supersymmetry.

The observed mixing angles among neutrino flavours are both large,
which is completely different from the quark sector.  This is a big
challenge in constructing a realistic model of SO(10) GUT, since it
describes quarks and leptons in unified way.  One simple possibility
has been proposed in the so-called ``lopsided''
models~\cite{Sato:1997hv,Albright:1998vf,Irges:1998ax}.  Mass matrices
for down quarks and leptons are arranged to have the lopsided
structure, {\it i.e.}, off-diagonal elements appear in a lopsided way
such that mixings of left-handed leptons are large while those of
left-handed down quarks are small, which give desired mixings
of quarks and leptons in the charged current.
There have been proposed so far many models to realize the
lopsided mass matrices~\cite{Barr:2000ka}.

Such mass matrices can be based on the Froggatt-Nielsen
(FN) mechanism~\cite{Froggatt:1978nt}.  It has been shown in
Ref.~\cite{Sato:1997hv} by using SU(5) GUT that the
observed mass hierarchies of quarks and charged leptons are explained
by their lopsided charges under some FN flavour symmetry, and the
large $\nu_\mu$-$\nu_\tau$ mixing angle in the atmospheric neutrinos
is a direct consequence of such lopsided charge assignment.  In this
letter we would like to realize this idea in the framework of SO(10)
GUT and show one possibility to yield the successful mass matrices
obtained in the lopsided SU(5) model.  We also discuss cosmological
implication of the model, in particular, possible scenarios of
leptogenesis in the inflationary Universe.

Masses of quarks and charged leptons approximately satisfy the following 
relations at the unification scale $M_{GUT} \simeq 2\times10^{16}$ GeV:
\begin{eqnarray}
  &&m_u : m_c : m_t \simeq \epsilon^4 : \epsilon^2 : 1~,
  \nonumber \\
  &&m_e : m_\mu : m_\tau \simeq m_d : m_s : m_b \simeq 
  \epsilon^3 : \epsilon : 1~,
  \label{eq:MR}
\end{eqnarray}
where $\epsilon \sim 1/16$.
One attractive attempt to understand mass hierarchies and mixings
among fermions consists in calling upon the
Froggatt-Nielsen (FN) mechanism~\cite{Froggatt:1978nt}.  
This mechanism can be based on U(1)$_{FN}$ flavor symmetry, 
which is broken by a vacuum expectation value (vev) of a gauge singlet field
$S_{FN}$ carrying the $U(1)_{FN}$ charge $- 1$.%
\footnote{We take here U(1)$_{FN}$ 
as the FN flavour symmetry by way of illustration.
This U(1)$_{FN}$ can be replaced by some discrete symmetry 
which is anomaly-free.}
Matter $f_i$ and Higgs $H$ superfields are introduced with
charges $Q_i$ ($Q_i \ge 0$) and 0, respectively.  
Then, the U(1)$_{FN}$ flavor symmetry
allows the following Yukawa terms 
\begin{eqnarray}
  W = c_{ij} \left( \frac{ S_{FN} }{ M_\ast } \right)^{Q_i +Q_j }
  ~ H ~ f_i f_j ~,
\end{eqnarray}
where $M_\ast$ denotes the fundamental cutoff scale of the theory
and we regard it as the gravitational scale.
By assuming that constant coefficients $c_{ij}$ are of order unity,
fermion mass matrices scale as 
$m_{ij} \propto \epsilon^{Q_i + Q_j}$ depending on 
the U(1)$_{FN}$ charges, where the parameter $\epsilon$ is 
defined by
\begin{eqnarray}
  \epsilon \equiv \frac{\vev{S_{FN}}}{M_\ast}~.
  \label{eq:defEP}
\end{eqnarray}
We set $\epsilon \sim 1/16$ from the mass relations shown in Eq.~(\ref{eq:MR}).

%
%%%%%%%%%%%%%%%%%%%%%%%%%%%%%%%%%%%%%%%%%%%%%%%%%%%%%%%%%%%%%%%%%%%
%%%% Table for matter and Higgs multiplets in SU(5)
%%%%%%%%%%%%%%%%%%%%%%%%%%%%%%%%%%%%%%%%%%%%%%%%%%%%%%%%%%%%%%%%%%%
\renewcommand{\arraystretch}{1.2}
\begin{table}[t]
  \begin{center}
    $\begin{array}[h]{| c | c c c | c c c | c c c |}\hline
      f_i & {\bf 10}_3 & {\bf 10}_2 & {\bf 10}_1 & 
      {\bf 5^\ast}_3 & {\bf 5^\ast}_2 & {\bf 5^\ast}_1 &
      {\bf 1}_3 & {\bf 1}_2 & {\bf 1}_1
      \\ \hline
      \mbox{U(1)}_{FN} & 0 & 1 & 2 & a & a & a+1 & b & c & d
      \\ \hline
    \end{array}$
    \caption{U(1)$_{FN}$ charge assignment in SU(5) GUT. 
      Here $a=0$ or $1$ and $0 \le b \le c \le d$.}
    \label{tab:SU5}
  \end{center}
\end{table}
%%%%%%%%%%%%%%%%%%%%%%%%%%%%%%%%%%%%%%%%%%%%%%%%%%%%%%%%%%%%%%%%%%%%
%%%%%%%%%%%%%%%%%%%%%%%%%%%%%%%%%%%%%%%%%%%%%%%%%%%%%%%%%%%%%%%%%%%%
The above FN mechanism can be used to construct lopsided mass matrices
for down quarks and leptons, which explain the fermion mass relations
in Eq.~(\ref{eq:MR}) as well as the observed large neutrino mixing
angles while keeping quark mixings small. First, we shall briefly
review this point by using SU(5) model, which have been discussed in
Refs.~\cite{Sato:1997hv,Buchmuller:1998zf}.

In SU(5) GUT, one family of quarks and leptons can be grouped into 
the SU(5) multiplets, ${\bf 10}$-plet, ${\bf 5^\ast}$-plet and 
${\bf 1}$-plet.
Their Yukawa interactions are given by
\begin{eqnarray}
  W =   
    h^u{}_{ij} H_u {\bf 10}_i {\bf 10}_j
  + h^d{}_{ij} H_d {\bf 10}_i {\bf 5^\ast}_j
  + h^D{}_{ij} H_u {\bf 1}_i {\bf 5^\ast}_j
  + h^N{}_{ij} S   {\bf 1}_i {\bf 1}_j
~,
\end{eqnarray}
where $H_u$ and $H_d$ denote Higgs fields of ${\bf 5}$-plet and ${\bf
  5^\ast}$-plet and their vevs are denoted by $v_2$ and $v_1$,
respectively.  Here we assumed that Majorana masses for right-handed
neutrinos come from the vev of the singlet Higgs $S$.  The
hierarchies in the Yukawa couplings $h^{u,d,D,N}$ are explained by the
FN mechanism.  In Table~\ref{tab:SU5} we show the U(1)$_{FN}$ charge
assignment for matter fields~\cite{Sato:1997hv,Buchmuller:1998zf}.
Then, we obtain mass matrices for quarks and charged leptons as
\begin{eqnarray}
  \label{eq:MSU5}
  m^u \eqn{=} v_2 ~h^u
  =
  v_2\left( 
    \begin{array}{c c c}
      \epsilon^4 & \epsilon^3 & \epsilon^2 \\
      \epsilon^3 & \epsilon^2 & \epsilon \\
      \epsilon^2 & \epsilon & 1 \\
    \end{array}
  \right),~~~
  m^e = (m^d)^T = v_1 ~ h^d
  =
  v_1~\epsilon^a \left( 
    \begin{array}{c c c}
      \epsilon^3 & \epsilon^2 & \epsilon^2 \\
      \epsilon^2 & \epsilon & \epsilon \\
      \epsilon & 1 & 1 \\
    \end{array} 
  \right)~.
\end{eqnarray}
It should be noted that every component in these mass matrices
contains a coefficient of order unity, which are implicitly assumed
here and hereafter.  It is found that the ``lopsided'' charge
assignment between ${\bf10}$-plets and ${\bf 5^\ast}$-plets leads to
the lopsided mass matrices for down quarks and charged leptons
(compared with the mass matrix for up quarks),
which is crucial for explaining the mass relations given in Eq.~(\ref{eq:MR}).
The mixing angles for quarks are found to be small.

On the other hand, the charge assignment in Table~\ref{tab:SU5}
can also explain large angles of neutrino flavour mixings.
Dirac and Majorana mass matrices for neutrinos, $m^D$ and $m^N$,
are given by
\begin{eqnarray}
  \label{eq:MnSU5}
  m^D = v_2 ~h^D = 
  v_2 ~ \epsilon^a \left( 
    \begin{array}{c c c}
      \epsilon^{d+1} & \epsilon^d & \epsilon^d \\
      \epsilon^{c+1} & \epsilon^c & \epsilon^c \\
      \epsilon^{b+1} & \epsilon^b & \epsilon^b \\
    \end{array}
  \right),
  ~~
  m^N = \vev{S} h^N  =
  \vev{S}\left( 
    \begin{array}{c c c}
      \epsilon^{2d} & 0 & 0 \\
      0 & \epsilon^{2c} & 0 \\
      0 & 0 & \epsilon^{2b} \\
    \end{array}
    \right),
\end{eqnarray}
where we have chosen a basis where $m^N$ is diagonal and real.
It is seen that $m^D$ takes also the lopsided form.
Then, the seesaw mechanism~\cite{SeeSaw} brings us the mass matrix 
for three light neutrinos as
\begin{eqnarray}
  \label{eq:MnuSU5}
  m^\nu = - (m^D)^T \frac{1}{m^N} m^D =
  \frac{v_2^2}{\vev{S}} ~\epsilon^{2a}~
  \left(
  \begin{array}{c c c}
      \epsilon^2 & \epsilon & \epsilon \\
      \epsilon    & 1 & 1 \\
      \epsilon    & 1 & 1 \\
    \end{array}
\right)~.
\end{eqnarray}
Notice that the dependence on the FN charges of right-handed neutrinos
drops off in this expression.  It is clearly seen that the angle of a
$\nu_\mu$-$\nu_\tau$ mixing is large, which is observed in the
atmospheric neutrino experiments.  This is a direct consequence of the
fact that ${\bf 5^\ast}$-plets in the second and third families have
the same FN charges in order to explain the mass hierarchies of down
quarks and charged
leptons~\cite{Sato:1997hv,Albright:1998vf,Irges:1998ax}.  Furthermore,
it have been shown in Ref.~\cite{Vissani:1998xg} that there is no
difficulty in this mass matrix to give the
large mixing angle in the solar neutrinos.

The above lopsided SU(5) model gives the successful mass matrices for
quarks and leptons, to the first approximation.  To explain the mass matrices
more precisely, we have to include effects of SU(5) breaking.
Otherwise, the unwanted SU(5) mass relations $m_e = m_d$ and $m_\mu =
m_s$ are obtained.  This can be achieved by introducing additional
Higgs fields.  However, this issue is beyond the scope of this
analysis.

In SO(10) GUT, one family of quarks and leptons can be simply grouped
into an irreducible spinor representation ({\bf \underline{16}}-plet).%
\footnote{ The representations with and without underline correspond
  to those under SO(10) and SU(5) gauge groups, respectively. }  
If it is the case,  quarks and leptons in each family possess
the same FN charge and all mass matrices of quarks and leptons
have the same hierarchical structure.  This means that we fail to 
explain the mass relations (\ref{eq:MR}).
To avoid this difficulty, we introduce 
three {\bf \underline{16}}-plets $\psi_i$ ($i=1,2,3$) and also one additional
{\bf \underline{10}}-plet $\eta$ as matter fields:
\begin{eqnarray}
  \psi_i = (~ {\bf 10}_i,~{\bf 5}^\ast{}_i,~{\bf 1}_i ~)~~~\mbox{and}~~~
  \eta   = (~ {\bf 5}_4,~{\bf 5}^\ast{}_4 ~)~,
\end{eqnarray}
where we have shown the decomposition under SU(5) group.
In Table~\ref{tab:Content} we show the FN charges for 
these matter superfields.  We determine the charges of $\psi_i$
from the mass hierarchy of up quarks and assign zero charge to $\eta$.
In the followings, we will exchange one combination of ${\bf 5}^\ast_i$
in $\psi_i$ for ${\bf 5}^\ast_4$ in $\eta$, which have been 
proposed in the different SO(10) model~\cite{Nomura:1998gm}.
By this exchange we can have the lopsided structure 
between ${\bf 10}$-plets and ${\bf 5^\ast}$-plets in three families.
%
%%%%%%%%%%%%%%%%%%%%%%%%%%%%%%%%%%%%%%%%%%%%%%%%%%%%%%%%%%%%%%%%%%%
%%%% Table for matter and Higgs multiplets
%%%%%%%%%%%%%%%%%%%%%%%%%%%%%%%%%%%%%%%%%%%%%%%%%%%%%%%%%%%%%%%%%%%
\renewcommand{\arraystretch}{1.2}
\begin{table}[t]
  \begin{center}
    $\begin{array}[h]{| c | c c c c | }\hline
      {f_i} & \psi_3 ({\bf \underline{16}}) 
      & \psi_2 ({\bf \underline{16}})
      & \psi_1 ({\bf \underline{16}})
      & \eta   ({\bf \underline{10}})
      \\ \hline
      \mbox{U(1)}_{FN} & 0 & 1 & 2 & 0 
      \\ \hline
    \end{array}$
    \caption{U(1)$_{FN}$ charge assignment in SO(10) GUT.}
    \label{tab:Content}
  \end{center}
\end{table}
%%%%%%%%%%%%%%%%%%%%%%%%%%%%%%%%%%%%%%%%%%%%%%%%%%%%%%%%%%%%%%%%%%%%
%%%%%%%%%%%%%%%%%%%%%%%%%%%%%%%%%%%%%%%%%%%%%%%%%%%%%%%%%%%%%%%%%%%%
%

We introduce here the following Higgs superfields
\begin{eqnarray}
  H_1 ~({\bf \underline{10}}),~
  H_2 ~({\bf \underline{10}}),~
  \Phi ~({\bf \underline{16}}),~
  \Phi^c ~({\bf \underline{16}^\ast}),~
  \Sigma ~({\bf \underline{16}})~
  \mbox{and}~
  \Sigma^c ~({\bf \underline{16}^\ast})~.
\end{eqnarray}
Two Higgs doublets which couple to down quarks and up quarks 
are assumed to be contained in $H_1$ and $H_2$, respectively.%
\footnote{ In the SO(10) models discussed in
  Refs.~\cite{Nomura:1998gm,Nomura:1999ty} one {\bf \underline{10}}-plet and one
  {\bf \underline{16}}-plet are introduced for the Higgs doublets.  The
  considering two {\bf \underline{10}}-plets for the weak Higgs
  doublets are a natural consequence of the model in six dimensions
  where SO(10) breaking is achieved by orbifold
  compactification~\cite{Asaka:2001eh,Hall:2001xr}.
  In this case, the mass splitting between 
  the weak doublet and the color triplet Higgs fields is 
  realized naturally. } 
We assume the vevs of these Higgs fields as follows:
\begin{eqnarray}
  \vev{H_1} = v_1,~ \vev{H_2} = v_2,~ 
  \vev{\Phi} = \vev{\Phi^c} = v_\Phi~~
  \mbox{and}~
  \vev{\Sigma} = \vev{\Sigma^c} = v_\Sigma~,
\end{eqnarray}
where $v_1^2 + v_2^2 = (174~\GeV)^2$ and $v_\Phi$ and $v_\Sigma$ are
of order of the unification scale $M_{GUT}$ which keep an SU(5) subgroup
unbroken.  All these Higgs fields carry zero charge under the
U(1)$_{FN}$ symmetry (see, however, the discussion in the footnote
4.).

Now we are at the point to discuss mass matrices for quarks and
leptons.  The Yukawa interactions we shall consider here
are given by the following superpotential:
\begin{eqnarray}
  W &=& \frac 12 h^d_{ij}~ H_1 ~\psi_i \psi_j ~
  + ~\frac 12 h^u_{ij}~ H_2 ~\psi_i \psi_j ~
  + ~\frac 12 h^n_{ij}~ \frac{ \Phi^c \Phi^c }{ M_\ast } ~ \psi_i \psi_j~
  \nonumber \\
  &&
  + ~ g^\eta_i ~ \Sigma ~ \psi_i \eta~
  + ~ g^d_i~ \frac{ \Sigma^c }{ M_\ast } ~H_1~ \psi_i \eta ~
  + ~ g^u_i~ \frac{ \Sigma^c }{ M_\ast } ~H_2~ \psi_i \eta ~,
  \label{eq:SP}
\end{eqnarray}
where $h^{d,u,n}$ and $g^{\eta,d,u}$ denote Yukawa coupling constants 
which are explained by the FN charges of matter fields.
In this equation, the first two terms give usual Dirac masses
for quarks and leptons, the third term gives Majorana masses
for right-handed neutrinos, and the rest three terms denote
mass mixings between matter {\bf \underline{16}}-plets and 
additional {\bf \underline{10}}-plet.

Let us first discuss the exchange of ${\bf 5}^\ast$-plets in
matter fields. The vev of the $\Sigma$ field induces
\begin{eqnarray}
  W ~=~ g_i^\eta ~v_\Sigma~{\bf 5}_4 ~ {\bf 5^\ast}_i~,
\end{eqnarray}
which gives a Dirac mass for one linear combination (denoted by ${\bf 5}^\ast_H$)
among three ${\bf 5^\ast}_i$ in $\psi_i$, while ${\bf 5^\ast}_4$ in 
$\eta$ is still massless~\cite{Nomura:1998gm}.
From the FN charge assignment in Table~\ref{tab:Content} 
the Yukawa couplings $g^\eta{}_i$ are given by 
$g^\eta{}_i = (~\epsilon^2,~\epsilon,~1~)$
with $\epsilon \sim 1/16$.
Note again that coefficients of order unity are implicitly assumed.
We find, then, the dominant component of ${\bf 5^\ast}_H$ is ${\bf 5^\ast}_3$:
\begin{eqnarray}
  {\bf 5^\ast}_H \simeq \epsilon^2 {\bf 5^\ast}_1
  + \epsilon {\bf 5^\ast}_2 + {\bf 5^\ast}_3 \simeq {\bf 5^\ast}_3~.
\end{eqnarray}
In the following analysis 
we will take ${\bf 5^\ast}_H = {\bf 5^\ast}_3$ for simplicity.
Since the mass of ${\bf 5^\ast}_H$ is estimated as
$g^\eta{}_3 v_\Sigma \simeq v_\Sigma \sim M_{GUT}$,
three families of quarks and leptons at low energies are given by
\begin{eqnarray}
  \begin{array}{c c c}
    {\bf 10}_1 (2), & {\bf 10}_2 (1), & {\bf 10}_3 (0), \\
    {\bf 5^\ast}_1 (2), & {\bf 5^\ast}_2 (1), & {\bf 5^\ast}_4 (0),\\
    {\bf 1}_1(2), & {\bf 1}_2 (1), & {\bf 1}_3(0).
  \end{array}
\end{eqnarray}
Notice that there appears no other massless matter field.
Here we have also shown the FN charge of each multiplet.
Even after the exchange of ${\bf 5^\ast}$-plets
the FN charge assignment does not have the lopsided structure.
As we will explain below, the Higgs field $\Sigma^c$ plays 
an important role to generate the lopsided mass matrices.

Up quarks obtain Dirac masses from the usual Yukawa term
$W = \frac12 h^u{}_{ij} H_2  \psi_i \psi_j$
in the superpotential~(\ref{eq:SP}) and 
its mass matrix takes the form
\begin{eqnarray}
  m^u{} = v_2 ~h^u{} = v_2~
  \left( 
    \begin{array}{c c c}
      \epsilon^4 & \epsilon^3 & \epsilon^2 \\
      \epsilon^3 & \epsilon^2 & \epsilon \\
      \epsilon^2 & \epsilon & 1 \\
    \end{array}
\right)~.
\end{eqnarray}
This is the same as in the SU(5) model (cf. Eq.~(\ref{eq:MSU5}))
and gives the approximate mass relations shown in Eq.~(\ref{eq:MR}).
On the other hand, down quarks and charged leptons receive masses
from the following terms in the superpotential (\ref{eq:SP})
\begin{eqnarray}
  W =\frac12 h^d{}_{ij}~H_1~ \psi_i\psi_j
  +
  g^d{}_i ~\frac{\Sigma^c}{M_\ast} ~ H_1 ~\psi_i \eta
  =  h^d{}_{ij}  \vev{H_1} {\bf 10}_i {\bf 5^\ast}_j 
  +  g^d{}_i ~\epsilon_\Sigma  \vev{H_1} {\bf 10}_i {\bf 5^\ast}_4 
  +\cdots~,
\end{eqnarray}
where $\epsilon_\Sigma \equiv v_\Sigma/M_\ast$.
Since ${\bf 5^\ast}_H (={\bf 5^\ast}_3)$ decouples from the 
low energy physics, we have the effective mass terms 
for down-quarks and charged leptons as
\begin{eqnarray}
  W =
  \vev{H_1}
  \left(~ {\bf 10}_1,~{\bf 10}_2,~{\bf 10}_3~\right)
  \left(
    \begin{array}{c c c}
      \epsilon^4 & \epsilon^2 \epsilon_\Sigma & \epsilon^3 \\  
      \epsilon^3 & \epsilon   \epsilon_\Sigma & \epsilon^2 \\
      \epsilon^2 &            \epsilon_\Sigma & \epsilon \\
    \end{array}
  \right)
  \left(
    \begin{array}{c}
      {\bf 5^\ast}_1 \\
      {\bf 5^\ast}_4 \\
      {\bf 5^\ast}_2 \\
    \end{array}
  \right)~.
\end{eqnarray}
Mass matrices for down quarks and charged leptons are
\begin{eqnarray}
  \label{eq:Ml}
  m^e = (m^d)^T = v_1 
    \left(
    \begin{array}{c c c}
      \epsilon^4 & \epsilon^2 \epsilon_\Sigma & \epsilon^3 \\  
      \epsilon^3 & \epsilon   \epsilon_\Sigma & \epsilon^2 \\
      \epsilon^2 &            \epsilon_\Sigma & \epsilon \\
    \end{array}
  \right)~,
\end{eqnarray}
which yields the following mass relations
\begin{eqnarray}
  m_d : m_s : m_b = m_e : m_\mu : m_\tau \simeq
  \epsilon^3 : \epsilon_\Sigma : 1~.
\end{eqnarray}
It should be noted that, when $\epsilon_\Sigma$ is equal to
$\epsilon$, the mass matrices in Eq.~(\ref{eq:Ml}) coincide with those
in the SU(5) lopsided model with $a=1$ (see Eq.~(\ref{eq:MSU5})).  In
the considering SO(10) model, the $\Sigma^c$ field plays partially the
same role as the FN field $S_{FN}$ only for ${\bf 5^\ast}_4$, and the
successful mass matrices for down quarks and leptons
are obtained when we arrange the vev of $\Sigma^c$ such that
$\epsilon_\Sigma \simeq \epsilon$.  
In this analysis we consider that ${\bf  5^\ast}_4$ belongs to 
the second family as in Refs.~\cite{Nomura:1998gm,Nomura:1999ty}.  
It can, however, belong to the third family 
as long as $\epsilon_\Sigma \simeq \epsilon$.
The ratio of vevs between Higgs doublets is estimated as
$\tan \beta = v_2/v_1 \simeq \epsilon ~m_t/m_b \sim 8$
at the unification scale.

We turn to discuss neutrino masses.  Dirac mass terms
are generated in the similar way to down quarks and charged leptons.
\begin{eqnarray}
  W \eqn{=} \frac12 h^u{}_{ij}~H_2 ~\psi_i \psi_j ~
  + ~ g^u{}_i ~\frac{\Sigma^c}{M_\ast}~H_2~ \psi_i \eta
  \nonumber \\
  \eqn{=} \vev{H_2} 
  \left(~ {\bf 1}_1,~{\bf 1}_2,~{\bf 1}_3~\right)
  \left(
    \begin{array}{c c c}
      \epsilon^4 & \epsilon^2 \epsilon_\Sigma & \epsilon^3 \\  
      \epsilon^3 & \epsilon   \epsilon_\Sigma & \epsilon^2 \\
      \epsilon^2 &            \epsilon_\Sigma & \epsilon \\
    \end{array}
  \right)
  \left(
    \begin{array}{c}
      {\bf 5^\ast}_1 \\
      {\bf 5^\ast}_4 \\
      {\bf 5^\ast}_2 \\
    \end{array}
  \right) + \cdots~,
\end{eqnarray}
and hence we have
\begin{eqnarray}
  m^{D}= v_2
    \left(
    \begin{array}{c c c}
      \epsilon^4 & \epsilon^2 \epsilon_\Sigma & \epsilon^3 \\  
      \epsilon^3 & \epsilon   \epsilon_\Sigma & \epsilon^2 \\
      \epsilon^2 &            \epsilon_\Sigma & \epsilon \\
    \end{array}
  \right)~.
\end{eqnarray}
It is seen that $m^D$ has the same lopsided structure
as $m^e$ and $(m^d)^T$. Majorana masses
for right-handed neutrinos are induced by the following terms,
$W = h^n{}_{ij} ~\frac{ \Phi^c \Phi^c }{ M_\ast } ~ \psi_i \psi_j$,
which leads to
\begin{eqnarray}
  \label{eq:MN}
  m^n = \frac{v_\Phi^2}{M_\ast}~h^n = \frac{ v_\Phi^2 }{ M_\ast }
  \left( 
    \begin{array}{c c c}
      \epsilon^4 & \epsilon^3 & \epsilon^2 \\
      \epsilon^3 & \epsilon^2 & \epsilon \\
      \epsilon^2 & \epsilon & 1 \\
    \end{array}
  \right)~.
\end{eqnarray}
Through the seesaw mechanism we obtain the following
Majorana mass matrix for the left-handed neutrinos
\begin{eqnarray}
  \label{eq:Mnu}
  m^{\nu} = 
  \frac{ v_2^2 M_\ast }{ v_\Phi^2 } 
  \left( 
    \begin{array}{c c c}
      \epsilon^4 & \epsilon^2 \epsilon_\Sigma & \epsilon^3 \\
      \epsilon^2 \epsilon_\Sigma & \epsilon_\Sigma^2 
      & \epsilon \epsilon_\Sigma \\
      \epsilon^3 & \epsilon \epsilon_\Sigma & \epsilon^2 \\
    \end{array}
  \right)~.  
\end{eqnarray}
Just as before, 
when we set $\epsilon_\Sigma \simeq \epsilon$,
we have the same structure as Eq.~(\ref{eq:MnuSU5}) 
in the lopsided SU(5) model.
In this case,  we can have naturally a large mixing angle for the 
$\nu_\mu$-$\nu_\tau$ atmospheric neutrino oscillation, 
and we have no difficulty explaining a large mixing angle
in solar neutrino oscillation.

We denote eigenvalues of the mass matrix (\ref{eq:Mnu})
by $m_i$ ($m_1 < m_2 < m_3$). Identifying $m_3$
with $\sqrt{\delta m_{\rm atm}^2} \simeq 5 
\times 10^{-2}~\eV$ and requiring $\epsilon_\Sigma \simeq \epsilon$
the vev $v_\Phi$ is estimated as
\begin{eqnarray}
  \label{eq:VPHI}
  v_\Phi ~\simeq~ \epsilon ~
  \frac{ v_2 M_\ast^{1/2} }{ (\delta m_{\rm atm}^2)^{1/4} }
  ~\simeq~ 2 \times 10^{15}~\GeV~,
\end{eqnarray}
where we have taken the cutoff scale $M_\ast$ as the gravitational
scale $M_{P}\simeq 2.4 \times 10^{18}$ GeV.  It is seen that the scale of
$v_\Phi$ is one order below
the unification scale.%
\footnote{If the Higgs field $\Phi^c$ carries the FN charge $+1$,
the vev $v_\Phi$ is estimated as $v_\Phi \simeq 4\times 10^{16}$ GeV
and is comparable to the unification scale.}
Further, we find from Eq.~(\ref{eq:MN}) 
Majorana masses $M_i$ ($M_1 < M_2 < M_3$) for right-handed neutrinos
$n_i$ as
\begin{eqnarray}
  \label{eq:MNi}
  M_1 \simeq 4 \times 10^{7}~\GeV,~
  M_2 \simeq 9 \times 10^{9}~\GeV~\mbox{and}~
  M_3 \simeq 2 \times 10^{12}~\GeV~.
\end{eqnarray}
Notice that, in the considering SO(10) model,
the FN charges for right-handed neutrinos are fixed by 
those for up quarks, and hence we can predict Majorana masses uniquely.%
\footnote{ Even if we consider the non-zero FN charge for the Higgs
  field $\Phi^c$ (see footnote 4), the prediction of Majorana masses
  $M_i$ does not change.}  This is completely different from the
SU(5) model where they are determined by 
the unknown FN charges $b$, $c$ and $d$ (see Table~\ref{tab:SU5}).
This point is crucial for considering possible scenarios
of leptogenesis later.

We have shown that the successful mass matrices
of quarks and leptons in the SU(5) model can be obtained
in the framework of SO(10) GUT 
when $\epsilon_\Sigma \simeq \epsilon \sim 1/16$.
This means the vev of the Higgs field $\Sigma^c$ 
should be comparable to the vev of the FN singlet field $S_{FN}$ as
\begin{eqnarray}
  \label{eq:VSIG}
  v_\Sigma \simeq \vev{S_{FN}} 
  \simeq \epsilon ~ M_\ast ~
  \simeq
  2 \times 10^{17}~\GeV ~,
\end{eqnarray}
where we have taken $M_\ast = M_P$.  It is found that there
should be a new physics scale one order above the unification scale.
In fact, the SO(10) GUT is realized only above this scale and we have
SU(5) as an unbroken group for the scale $v_\Sigma \ge \mu \ge
M_{GUT}$.  If the origin of the vev $\vev{S_{FN}}$ is associated with
the SO(10) breaking, this GUT breaking pattern might
answer the required condition $\epsilon_\Sigma \simeq \epsilon$,
although we have to tune the scale correctly.%
\footnote{ One explanation for 
$\epsilon_\Sigma \simeq \epsilon \simeq 1/16$ might be obtained
by embedding the considering SO(10) GUT model in the higher dimensional 
theory.  We expect the cutoff scale $M_\ast$ is given by
the $(4+d)$-dimensional Planck scale, and hence
$M_\ast = (M_P^2 M_C^d)^{1/(d+2)}$ where $M_C$ denotes the
compactification scale of the extra $d$-dimensional space.
We consider $M_C$ is comparable to the unification scale.
When the SO(10) GUT is embedded in 6-dimensions ($d=2$), 
we find $M_\ast \simeq 2 \times 10^{17}$ GeV, and then
$v_\Sigma \simeq \vev{S_{FN}} \simeq 1 \times 10^{16}$ GeV.
Moreover, when the FN charge for the Higgs field $\Phi^c$
is $+1$ (see the footnote 4), $v_\Phi$ is also estimated 
to be $1 \times 10^{16}~\GeV$.
This suggests that the SO(10) GUT in 6-dimensions
might realize the successful lopsided mass matrices by the vevs
$\vev{S_{FN}}$, $v_\Sigma$ and $v_\Phi$ which 
are all of order of the unification scale (or $M_C$).
It is very interesting to note that the attractive SO(10) GUT models 
have recently been constructed in 
6-dimensions~\cite{Asaka:2001eh,Hall:2001xr}.
(See also the footnote 3.) }

Finally, we would like to discuss cosmological implication of the
model, in particular, implication in ``leptogenesis''~\cite{Fukugita:1986hr}.
Non-equilibrium decays of right-handed neutrinos $n_i$ 
give an attractive mechanism to generate dynamically
the observed baryon asymmetry in the present Universe.
This is because these decays can generate a lepton number 
in the early Universe,
which is partially converted into a baryon number through the 
electroweak sphaleron processes~\cite{Kuzmin:1985mm}.
The CP asymmetry by the $n_i$ decay can be expressed by the parameter 
$\epsilon_i$ and is estimated as~\cite{Covi:1996wh}
\begin{eqnarray}
  \label{eq:EPi}
  \epsilon_i 
  &\equiv& 
  \frac{ \Gamma ( n_i \rightarrow \ell + H_u ) - 
    \Gamma (n_i \rightarrow \ell^\dagger + H_u^\dagger ) }
  { \Gamma ( n_i \rightarrow \ell + H_u ) + 
    \Gamma (n_i \rightarrow \ell^\dagger + H_u^\dagger ) }
  \nonumber \\
  &=&
  - \frac{1}{8\pi v_2^2} 
  \frac{ 1 }{ ( m^D m^D{}^\dagger )_{ii} }
  \sum_{j\neq i} 
  \mbox{Im} \left[ \left\{ (m^D m^D{}^\dagger )_{ij} \right\}^2 \right]
  ~ f \left( \frac{ M_j^2 }{ M_i^2 } \right) ~,
\end{eqnarray}
where $n_i$, $\ell$ and $H_u$ denote here scalar or fermionic components
of corresponding superfields ($\ell$ are lepton doublets at low energies) 
and
\begin{eqnarray}
  f (x) = \sqrt{x} \ln \left( 1 + \frac{1}{x} \right) +
   \frac{ 2 \sqrt{x} }{ x - 1 }~.
\end{eqnarray}

The SO(10) model described above can predict Majorana masses for
right-handed neutrinos as shown in Eq.~(\ref{eq:MNi}) and their mass
ratios are determined by those of up quarks, which is completely
different from the SU(5) model.  This suggests possible scenarios of
leptogenesis are restricted.  The CP asymmetry $\epsilon_1$ of the
lightest right-handed neutrino $n_1$ is estimated from
Eq.~(\ref{eq:EPi}) as $\abs{\epsilon_1} \simeq 1 \times 10^{-8}$,
where we have assumed the CP-violating phase of order unity.  It have
been shown in Ref.~\cite{Buchmuller:1998zf} that the conventional
thermal leptogenesis~\cite{Buchmuller:2000as} does not work since
$\epsilon_1$ is too small to account for the present baryon asymmetry.

We, then, consider non-thermal leptogenesis via inflaton
decay~\cite{Lazarides:wu,Kumekawa:1994gx,Lazarides:1996dv,Giudice:1999fb,Asaka:1999yd}, where 
right-handed neutrinos are produced non-thermally in decays 
of inflaton $\varphi$.
The baryon asymmetry (the ratio of baryon number density $n_B$ 
to the entropy density $s$) induced by $n_1$ is given by%
\footnote{In this analysis, we neglect the sign of the produced baryon
  asymmetry.}
\begin{eqnarray}
  \frac{n_B}{s} \simeq 0.5 ~Br_1~
  \abs{\epsilon_1} ~\frac{T_R}{M_{\varphi}}~,
\end{eqnarray}
where $M_{\varphi}$, $T_R$ and $Br_1$ denote the inflaton mass,
the reheating temperature, and the branching ratio 
of $\varphi \rightarrow n_1+n_1$, respectively.
In order to ensure the non-thermal production of $n_1$
we assume $M_{\varphi} > 2 M_1$ and $T_R \lesssim M_1$.
In the above model we estimate~as
\begin{eqnarray}
  \frac{n_B}{s} \sim
  6 \times 10^{-10}~ Br_1~
  \left( \frac{ T_R }{10^7~\GeV} \right)
  \left( \frac{ 2 M_1 }{ M_\varphi } \right)~,
\end{eqnarray}
which should be compared with the observation $(n_B/s)_{\rm OBS}
\simeq (0.1-1) \times 10^{-10}$.  It is found that the successful
leptogenesis is available only for the inflation models which give
\begin{eqnarray}
  \label{eq:PR1}
  M_{\varphi} \sim 10^9~\GeV,~~ T_R \sim 10^7~\GeV ~~\mbox{and}~~
  Br_1\sim 1~.  
\end{eqnarray}

Further, as recently proposed in Ref.~\cite{Asaka:2002zu},
decays of heavier right-handed neutrinos $n_2$ and $n_3$ can be 
a dominant source of the present baryon asymmetry.  
It is usually considered that decays of the lightest right-handed neutrino
are responsible to $(n_B/s)_{\rm OBS}$, although
decays of heavier ones also induce lepton asymmetry.
This is because after the decays of $n_2$ and $n_3$ 
the lightest right-handed neutrino can remain in thermal equilibrium
and wash out the lepton asymmetry from $n_2$ and $n_3$.
This is true for the conventional thermal leptogenesis.  However, 
in the non-thermal leptogenesis scenarios,
as shown in Ref.~\cite{Asaka:2002zu}, such dangerous wash-out effects
can be killed just requiring that the reheating temperature is 
$T_R \lesssim M_1$.  
We shall illustrate this idea in the described model, especially,
leptogenesis by decays of the next-to-lightest right-handed neutrino
$n_2$.  The successful scenario requires that $M_\varphi > 2 M_2$ and
also $T_R \lesssim M_1$, which means that both $n_1$ and $n_2$ are
produced non-thermally in inflaton decays and induce the lepton
asymmetry.  We find the CP asymmetry for $n_2$ 
is $\abs{\epsilon_2} \simeq 2 \times 10^{-6}$ and
\begin{eqnarray}
  \frac{n_B}{s} \sim
  \left[ 3 \times 10^{-12}~ Br_1 + 
    6 \times 10^{-10} ~Br_2 \right]
  \left( \frac{ T_R }{10^7~\GeV} \right)
  \left( \frac{ 2 M_2 }{ M_\varphi } \right)~.
\end{eqnarray}
This equation shows that the dominant contribution to the baryon
asymmetry comes from the decays of $n_2$ rather than the lightest one $n_1$
unless $Br_2 \ll Br_1$.  In this case, the successful leptogenesis
requires 
\begin{eqnarray}
  \label{eq:PR2}
  M_{\varphi} \sim 10^{11}~\GeV, ~~T_R \sim 10^7~\GeV~~\mbox{and}~~
  Br_2 \sim 1~.
\end{eqnarray}

We have seen that possible scenarios of leptogenesis are restricted in
the SO(10) model even for the non-thermal leptogenesis via inflaton
decays, and the present baryon asymmetry can suggest parameters of
inflation models as shown in Eqs.~(\ref{eq:PR1}) and (\ref{eq:PR2}).
It is quite interesting to observe that the SO(10) model described
above provides naturally the hybrid inflation and, moreover, the
values given in Eq.~(\ref{eq:PR2}) are just predicted
by this inflation.%
\footnote{See Refs.~\cite{Lazarides:1996dv,Asaka:1999yd} for the
  similar discussion in the different models.}
Let us write the superpotential which gives non-zero vev for the Higgs
fields $\Phi$ and $\Phi^c$ as follows;
\begin{eqnarray}
  W = \lambda X ( \Phi \Phi^c - v_\Phi^2 )~,
\end{eqnarray}
where $\lambda$ is the coupling constant.
This is nothing but the superpotential for the supersymmetric hybrid
inflation~\cite{Copeland:1994vg}.%
\footnote{ Supergravity effects are potentially dangerous since they
  disturb the slow-roll motion of inflation.  These effects are
  induced from the nonrenormalizable interaction in the K\"ahler
  potential $K = (k/4) \abs{X}^4/M_\ast^2$.  For the successful
  inflation the coupling $k$ should be negative and also $\abs{k}
  \lesssim 10^{-3}$ when $\lambda \sim 10^{-4}$.  We assume it by hand
  since smallness of couplings in the K\"aher potential cannot be
  explained by the FN mechanism. }
The value of $v_\Phi \simeq 2 \times 10^{15}$ GeV (see
Eq.~(\ref{eq:VPHI})), which is suggested from the atmospheric neutrino
oscillation, gives the coupling constant of $\lambda \sim 10^{-4}$ in
order to explain the COBE normalization of the cosmic density
fluctuations.%
\footnote{The coupling constant $\lambda$ of $10^{-4}$ can be
  explained by the FN mechanism with the FN charge $+3$ for $X$.  } 
(See the detailed analysis of the hybrid inflation in
Ref.~\cite{Asaka:1999yd}.)  The inflaton mass $M_\varphi$ is, then,
estimated as $M_\varphi = \sqrt{2} \lambda v_\Phi \sim 10^{11}$ GeV,
which is just the required value in Eq.~(\ref{eq:PR2}) for the
successful leptogenesis.  The reheating of inflation takes place via
decays through the interactions $W = \frac 12 h^n_{ij}~ \frac{ \Phi^c
  \Phi^c }{ M_\ast } ~ \psi_i \psi_j$ in Eq.~(\ref{eq:SP}).
Therefore, the inflaton decays mainly into pairs of right-handed
(s)neutrinos and their partial widths are proportional to $M_i^2$.%
\footnote{Although the inflaton decays also through interactions in
  the K\"ahler potential, they give negligible corrections to the
  total width. }
We find from Eq.~(\ref{eq:MNi}) that $Br_2 \simeq 1$ is ensured for
the inflaton with $M_\varphi \sim 10^{11}$ GeV.  Further, the
reheating temperature is estimated as $T_R \sim 10^7$ GeV.  With such
low reheating temperatures, we can avoid the cosmological gravitino
problem~\cite{Khlopov:pf}.%
\footnote{ We can neglect gravitinos produced non-thermally at the
  preheating epoch~\cite{Kallosh:1999jj} since the coupling $\lambda$
  is sufficiently small.  Similarly, production of right-handed
  neutrinos by the preheating~\cite{Giudice:1999fb} can also be
  neglected.  }
Therefore, the hybrid inflation provided by the model itself offers the
successful non-thermal leptogenesis by decays of the next-to-lightest
right-handed neutrinos.  The observed baryon asymmetry in the present
Universe is explained by just the neutrino mass scale suggested by the
atmospheric neutrino experiments.

~

\noindent 
The author would like to thank W.~Buchm\"uller for useful discussions.

%%%%%%%%%%%%%%%%%%%%%%%%%%%%%%%%%%%%%%%%%%%%%%%%%%%%%%%%%%%%%%%%%%%
%%%%%%%%%%%%%%%%%%%%%%%%%%%%%%%%%%%%%%%%%%%%%%%%%%%%%%%%%%%%%%%%%%%

%%%%%%%%%%%%%%%%%%%%%%%%%%%%%%%%%%%%%%%%%%%%%%%%%%%%%%%%%%%%%%%%%%%
%%%%%%%%%%%%%%%%%%%%%%%%%%%%%%%%%%%%%%%%%%%%%%%%%%%%%%%%%%%%%%%%%%%
\end{document}